\documentclass[12pt]{article}
\usepackage{amsmath,amsthm,amscd,amsfonts,amssymb}
\usepackage{latexsym}
\usepackage[usenames]{color}

\newcommand{\NN}{\mathbb N}

\newcommand{\RR}{\mathbb R}
\newcommand{\TT}{\mathbb T}
\newcommand{\ZZ}{\mathbb Z}

\newcommand{\hh}{\mathcal H}

\newcommand{\sS}{\mathcal S}

\newtheorem{thm}{Theorem}[section]
\newtheorem{lemma}[thm]{Lemma}
\newtheorem{cor}[thm]{Corollary}
\newtheorem{prop}[thm]{Proposition}
\newtheorem{defn}[thm]{Definition}

\def\half{\frac{1}{2}}

\def\pf{{\noindent \bf Proof: }}
\def\vlam{(V+\lambda)^{-1}}

\def\mvlam{(V+\lambda)^{-m}}
\def\mhlam{(H+\lambda)^{-m}}
\def\hmvlam{(V+\lambda)^{-\frac{m}{2}}}

\begin{document}

\title{Szeg\"o limit Theorem on the Lattice}

\author{Jitendriya Swain \\
Department of Mathematics \\
IIT, Guwahati \\
e-mail : jitumath@iitg.ernet.in \\
  and \\ 
M Krishna \\ 
The Institute of Mathematical Sciences \\
Taramani, Chennai 600113 \\
e-mail: krishna@imsc.res.in}

\date{\today}

\maketitle

\begin{abstract} In this paper, we prove a Szeg\"{o} type limit theorem
on $\ell^2(\ZZ^d)$.  We consider
operators of the form $H=\Delta+V$, $V$ multiplication by a positive sequence 
$\{V(n), n \in \ZZ^d\}$ with $V(n) \rightarrow \infty, |n| \rightarrow \infty
$
on $\ell^2(\ZZ^d)$ and
$\pi_{\lambda}$  the orthogonal projection of $\ell^2(\mathbb{Z}^d)$ on to
the space of eigenfunctions of $H$ with eigenvalues $\leq \lambda$.
We take $B$ to be a pseudo difference operator of order zero with symbol 
$b(x,n), (x,n) \in \TT^d\times \ZZ^d$  
and show that for nice functions $f$ 
$$
\lim_{\lambda \rightarrow \infty} Tr(f(\pi_\lambda B\pi_\lambda))/Tr(\pi_\lambda) = 
\lim_{\lambda \rightarrow \infty} \frac{1}{(2\pi)^d} \frac{\sum_{V(n) \leq \lambda} \int_{\TT^d} f(b(x,n)) ~ dx}{\sum_{V(n)\leq\lambda} 1 }.
$$
\end{abstract}

\vspace{-.5cm}

\section{Introduction}

In this paper we show a Szeg\"{o} type theorem on the lattice and give some application to
random operators.

The classical theorem of Szeg\"{o} is stated as follows: Let $P_n$ be the orthogonal projection of $L^2[0,2\pi]$ onto the linear subspace spanned by the functions 
$\{e^{im\theta}:0\leq m\leq n;0\leq\theta<2\pi\}$. For a positive function 
$f\in \mathcal{C}^{1+\alpha}[0,2\pi],\alpha>0$ the operator $T_f$ defined by the operator of 
multiplication by the function $f$ on $L^2[0,2\pi]$ the following result holds
$$
\lim_{n\to\infty}\frac{1}{n+1}\log \det P_nT_f P_n=\frac{1}{2\pi}\int_0^{2\pi}\log f(\theta)d\theta.
$$
The above result is well known as Szeg\"{o} limit theorem. We refer to \cite{sze1,sze2} for details 
and related results. In fact, Szeg\"{o} limit theorem is a special case of a more general result 
proved by Szeg\"{o} (see \cite{sze2}) in section $5.3$ as follows. Let $f$ be a bounded, real 
valued integrable function and $\{\lambda_i^n\}_{i=1}^n$ be the eigenvalues of $P_nT_fP_n$. 
Then for any continuous function $F$ on $[\inf f,\sup f]$ it was proved in 
(see \cite{sze2}, sect. 5.3) that
$$
\lim_{n\to\infty}\frac{1}{n}\sum_{i=1}^{n}F(\lambda_i^n)=\frac{1}{2\pi}\int_0^{2\pi}F( f(\theta))d\theta.
$$
Notice that the left hand side here can be seen to be the limit of 
$$
Tr(F(P_nT_fP_n))/Tr(P_n)
$$ 
and that $e^{im\theta}$ is an eigenfunction of $\Delta=-\frac{d^2}{dx^2}$, so, one can view the 
above results on $L^2[0,2\pi]$ as a special cases of Szeg\"{o} limit theorem for the 
Laplace-Beltrami operator or more generally one can consider such results for pseudo 
differential operators on compact manifolds.

In \cite{zel}, Zelditch considered a Schr\"{o}dinger operator on $\mathbb{R}^n$ of the form $H=-\frac{1}{2}\Delta+V$, where $V$ is a smooth positive function which grows like $V_0|x|^k,~k>0$. 
To establish a Szeg\"{o} type theorem, as we can see from the above,  we need to consider ratios of 
distribution functions associated to different measures and compare their behaviour asymptotically.

Such limits are computed using Tauberian theorems where some transforms of these measures are 
considered and limits taken for such transforms. While Zelditch \cite{zel} used the Laplace 
transform (via Karamata's Tauberian theorem (\cite{wid},p-192), Robert \cite{rob} suggested the 
use of Stieltjes transform (via Keldysh Tauberian theorem\cite{kel}). The application of 
Keldysh theorem requires one of the measures $\mu$ or $\nu$ to be absolutely continuous. 
We don't have this feature in our problem, stated below, so we use the Tauberian theorem of 
Grishin-Poedintseva theorem \ref{gp} (see \cite{gri}) in combination with a theorem 
of Laptev-Safarov theorem \ref{ls} (see \cite{LapSaff}) that obtains some error estimates
to prove our main theorem (Theorem \ref{szego}).

There is an extensive work on the Szeg\"o's theorem associated with orthogonal polynomials 
in $L^2(\TT, d\mu)$ with $\mu$ some probability measure on $\TT$, we refer to the monumental
work of Barry Simon \cite{sim} for the details.  

We however concentrate on higher dimensions where not much is known and to our knowledge our
results are new in the lattice case.

We consider operators of the form
\begin{equation}\label{eqn0}
H = \Delta + V 
\end{equation}
on $\ell^2(\ZZ^d)$, where $\Delta$
is the positive operator $(\Delta u)(n) = \sum_{|n-j|=1} u(j) + 2d u(n)$.
We take $V$ is multiplication by a positive sequence 
\begin{equation}\label{eqn1}
V(n) = \begin{cases}  1 \\  |n|^k \end{cases} , ~~~ k > 0
\end{equation}
we chose the value of $V(n)$ to be $1$ at the origin to make $V$ strictly positive.  

Then $H$ is positive and has discrete spectrum, which is seen by noting
that $(H-i)^{-1}$ is compact in view of the choice of $V$.  We denote
the spectral projection of $H$ by $E_H()$ and set 
$\pi_\lambda = E_H((0, \lambda])$.  Then clearly $\pi_\lambda$ is finite
rank for each $\lambda$.

For a bounded self-adjoint operator $B$ we set  
$K =  \bigcup_{0 \leq t \leq 1} \sigma(B) \subset \RR$,  
$L^2(\TT^d) = L^2(\TT^d, \frac{dx}{(2\pi)^d})$.  Then our main theorems are the following.

\begin{thm}\label{szego}
Let $H$ and $V$ be as in equation (\ref{eqn0}, \ref{eqn1}). 
Let $b$ be a bounded real valued measurable function on $\TT^d$,  
let $M_b$ be the operator of multiplication by $b$ on $L^2(\TT^d)$ and $B$  
its unitary equivalent on $\ell^2(\ZZ^d)$ under the Fourier Series.
Assume that there is a $0 < \kappa < 1$ such that $H^{-\kappa}[H, B]$ is bounded.
Then for all $f \in C(K)$, we have  
\begin{equation}
\lim_{\lambda \rightarrow \infty} \frac{Tr\left( f(\pi_\lambda B \pi_\lambda)\right) }{Tr(\pi_\lambda)} 
=\frac{1}{(2\pi)^d}\int f(b(x)) ~ dx  .
\end{equation}
\end{thm}

We recollect some facts on toroidal symbols from Rhuzanski-Turunen \cite{mic} below. A  
 linear operator 
$A$ on $L^2(\TT^d)$ associated with symbols $\sigma(x, n), (x, n) \in \TT^d\times \ZZ^d$, (the 
reader should note that the lattice variable $\xi$ appearing in \cite{mic} should be replaced
by $m, n$ etc in  our notation)  is defined by
\begin{equation}\label{symbol}
(A\phi)(x) = \sum_{n\in\ZZ^d} \frac{1}{(2\pi)^d} \int e^{i(x-y)n} \sigma(x, n) \phi(x) ~ dx 
\end{equation}
where $\phi \in C^\infty(\TT^d)$ and the symbol $\sigma \in C^\infty(\TT^d\times \ZZ^d)$, 
($a \in C^\infty(\TT^d\times \ZZ^d)$ means $a(\cdot, n) \in C^\infty(\TT^d)$ 
for all $n \in \ZZ^d$).  Then $A$ extends to a bounded linear operator and via the unitary
isomorphism implemented by the Fourier series (call it $U^*)$ 
between $L^2(\TT^d)$ and $\ell^2(\ZZ^d)$,
gives also a  bounded operator on $\ell^2(\ZZ^d)$.  Thus associated to every symbol 
$\sigma \in C^\infty(\TT^d\times \ZZ^d)$, there is a bounded operator on 
$\ell^2(\ZZ^d)$ and such an operator
is self-adjoint whenever $\sigma$ is real valued.  We will say
that $b(x,n)$ is the symbol of a bounded linear operator $B$ on $\ell^2(\ZZ^d)$
to mean that the equation (\ref{symbol}) is valid by setting $A=UBU^*, \sigma(x,n) = b(x,n)$
and $B$ is an appropriate bounded linear extension in $\ell^2(\ZZ^d)$.  
The difference operator
$\Delta_{n_j}$ is given by $(\Delta_{n_j}\phi)(m) = \phi(m+e_j) - \phi(m)$, $e_j$ being the unit
vector in the $j^{\mathrm{th}}$ direction in $\ZZ^d$ and acts on symbols in the second
variable.  Denoting $\NN_0 = \NN \cup \{0\}$,
and a multi index $\alpha = (\alpha_1, \dots, \alpha_d) \geq 0$ 
to mean $\alpha_j \in \NN_0, j=1,\dots, d$. Then the difference operator 
$\Delta_{n}^\alpha = \Delta_{n_1}^{\alpha_1}\Delta_{n_2}^{\alpha_2}\cdots \Delta_{n_d}^{\alpha_d}$.
Let $\langle n\rangle = (1 + |n|^2)^{\half}$.  The class of rapidly 
decreasing sequences is given by 
$$
\sS(\ZZ^d) =  \{\phi(n) : \forall ~ M \in \NN, |\phi(n)| \leq C_{\phi, M} \langle n\rangle^{-M}\}.
$$
Given these definitions, for any $m \in \RR, 0\leq \rho, \delta \leq 1$, 
the {\it toroidal symbol class} $S_{\rho, \delta}^m(\TT^d\times \ZZ^d)$
is defined as all $\sigma \in C^\infty(\TT^d\times \ZZ^d)$ such that   
$$
|\Delta_n^\alpha D_x^\beta \sigma(x,n) | \leq C_{\sigma\alpha \beta m} \langle n\rangle^{m-\rho|\alpha|+\delta|\beta|}, ~
\forall (x,n) \in \TT^d\times \ZZ^d ~ \mathrm{and}~ \alpha, \beta \in \NN_0^d.
$$ 
The class $S_{1,0}^0$ is denoted simply by $S_{1,0}$.
Let us define a subclass of symbols where all the derivatives in $x$ also have uniform bounds.
$$
S_{1,0, \infty}^m(\TT^d\times\ZZ^d) = 
\{\sigma \in S_{1,0}^m(\TT^d\times\ZZ^d) : C_{\sigma \alpha \beta m} ~ \mathrm{are ~ independent ~ of} ~ \beta\}. 
$$
So we denote $S_{1,0,\infty}^0$ by $S_{1,0,\infty}$.

Given this framework, we have our next theorem.

\begin{thm}\label{szego2}
Let $H$ and $V$ be as in equation (\ref{eqn0}, \ref{eqn1}). 
Consider a real valued $b \in S_{1,0, \infty}(\TT^d\times \ZZ^d)$ and let $B$ be
the associated bounded self-adjoint operator on $\ell^2(\ZZ^d)$.   
Assume that there is a $0 < \kappa < 1$ such that $H^{-\kappa}[H, B]$ is bounded.
Then for all $f \in C(K)$, we have  
\begin{equation}
\lim_{\lambda \rightarrow \infty} \frac{Tr\left( f(\pi_\lambda B \pi_\lambda)\right) }{Tr(\pi_\lambda)} 
=\lim_{\lambda \rightarrow \infty} \frac{ \frac{1}{(2\pi)^d} \int  \sum_{V(n) \leq \lambda } f(b(x,n)) ~ dx}{\sum_{V(n) \leq \lambda } 1 }  .
\end{equation}
\end{thm}

The function $\cos(x + \gamma_n)$ with $\gamma_n \rightarrow  0$ as $|n| \rightarrow \infty$
is in $S_{1,0,\infty}(\TT^d\times \ZZ^d$ for example, so the class is non-empty.
\section{The Proofs:}

We denoted by $\#S$ the cardinality of the set $S$ in the following.  Consider $H, V$
as in equations (\ref{eqn0}, \ref{eqn1}). 

Then for $\lambda > 0$ the operator $H+\lambda$ is also positive.  $H$ has discrete
spectrum and so the bounded operator $(H + \lambda)^{-1}$ is compact.  By taking
proper power $k$ we can make it trace class, so let us for the sake of simplicity 
assume $k > d$ so this operator is trace class.

Given this if we consider the operator $V$ and $(V +\lambda)$, then
these two are operators of multiplication on $\ell^2(\ZZ^d)$ , they both have discrete
spectrum and are positive so we also have , the our choice of $k$, that
$(V + \lambda)^{-m}$ and $\mhlam$ are  also trace class for some $m \in \NN$..

\begin{lemma}\label{lem1}
Consider $V$ and $H$ self-adjoint operators as given in equations (\ref{eqn1},\ref{eqn0}). Then
for $m \in \NN$ for which $\mvlam$ is trace class, $\mhlam$ is also trace class and  we have, 
$$
\vline ~ \frac{Tr((H + \lambda)^{-m})}{Tr((V+\lambda)^{-m})} - 1 \vline \rightarrow 0
$$
as $\lambda \rightarrow \infty$.
\end{lemma}
\pf Since $\Delta$ is bounded and $\vlam$ is bounded and positive 
we can write
$$
(H + \lambda) = (V +\lambda)^\half ( (V+\lambda)^{-\half} \Delta (V+\lambda)^{-\half} + 1 ) (V+\lambda)^{\half}.
$$
Since $K_\lambda = (V+\lambda)^{-\half }\Delta (V+\lambda)^{-\half}$ is bounded and has
norm smaller than 1 for large $\lambda$ , we have 
expansion
\begin{equation}\label{expand}
\mhlam = \mvlam + \hmvlam \left( (1+K_\lambda)^{-m} - 1)  \right) \hmvlam.
\end{equation}
This equality shows that $\mhlam$ is trace class whenever $\mvlam$ is trace class, so
we take trace on both sides of the above equation. We then use the 
property of trace, the inequality  
 $|Tr(B C B) | \leq \|C\| Tr(B^2)$, when $B$ is positive trace class
and $C$ is bounded and that $(1 + K_\lambda)^{-1}$ is bounded by 1, by the positivity of $K_\lambda$, 
to get
\begin{equation}
\begin{split}
&\vline ~ Tr(\mhlam) - Tr(\mvlam) ~ \vline \\
 & = \vline   Tr\left(\hmvlam \left( (1+K_\lambda)^{-m} - 1)  \right) \hmvlam\right)\vline \\ 
 & \leq Tr(\mvlam) \|\left( (1+K_\lambda)^{-m} - 1)  \right)\| \leq m \|K_\lambda\| Tr(\mvlam) \\
& \leq Tr(\mvlam) m \|\Delta\| \|\vlam\|^{m}, 
\end{split}
\end{equation}
Therefore,
$$
\vline ~ \frac{Tr(\mhlam)}{Tr(\mvlam)} - 1 ~ \vline \leq 4 d m \|\vlam\|^m 
$$
which gives the lemma as $\lambda$ goes to $\infty$. \qed

Let $E_A$ denotes the projection value spectral measure of  $A$.
Denote the distribution functions
of the measures  $Tr(E_H(\cdot))$ and $Tr(E_V(\cdot))$ respectively by 
$\phi_H$ and $\phi_V$.  Then
we have
$$
\phi_H(\lambda) = Tr(\pi_\lambda), ~~ \phi_V(\lambda) = \#\{ n : V(n) \in (0, \lambda]\}. 
$$

Then Lemma \ref{lem1} immediately gives us the Weyl formula for
the functions $Tr(\pi_\lambda)$ as a corollary, where we denote by $[r]$ the largest integer
smaller than or equal to $r$. 

\begin{cor}\label{cor1}
Consider $V$ and $H$ self-adjoint operators as given in equations (\ref{eqn0},\ref{eqn1}). 
We have the following asymptotics :
\begin{enumerate}
\item $\phi_V$ is multiplicatively continuous.
\item 
$$
Tr (\pi_\lambda) - \#\{n : V(n) \in (0, \lambda] \} \approx o(\#\{n : V(n) \in (0, \lambda] \}) , ~ \mathrm{as} ~ \lambda \rightarrow \infty.
$$
\item
$$
Tr(\pi_\lambda) = 2^d [\lambda]^{\frac{d}{k}} + o(\lambda^{\frac{d}{k}}) , ~ \mathrm{as} ~ \lambda \rightarrow \infty.
$$
\item 
$$
\sup_{\mu \leq \lambda} (Tr(\pi_{\mu+r}) - Tr(\pi_\mu)) \leq Tr(\pi_\lambda) \left( \frac{d}{k} \frac{[r]+1}{[\lambda]} + O(\frac{1}{[\lambda]})\right),~ \mathrm{as} ~ \lambda \rightarrow \infty.  
$$
\end{enumerate}
\end{cor}
\pf 
(1) The function $\phi_V$ is given by 
\begin{equation}\label{eqn6}
\begin{split}
\phi_V(\lambda) &= \#\{ n : V(n) \leq \lambda\} = \#\{n : |n|^k \leq \lambda\} \\ &= \#\{n : |n| \leq [\lambda]^{1/k}\} = (2[\lambda]^{\frac{1}{k}} + 1)^d.
\end{split}
\end{equation}
Therefore clearly $\lim_{\lambda \rightarrow \infty } \lim_{\tau \rightarrow 1} \phi_V(\tau \lambda) /\phi_V(\lambda) = 1$.
On the other hand, using the notation $(r)$ for the fractional part of $r$, we see from equation
(\ref{eqn6}) that
\begin{equation}\begin{split}
\frac{\phi_V(\tau\lambda)}{\phi_V(\lambda)} &= \frac{(2[\tau \lambda]^{\frac{1}{k}} + 1)^d}{(2[\lambda]^{\frac{1}{k}} + 1)^d} = \frac{(2(\tau \lambda - (\tau \lambda))^{\frac{1}{k}} + 1)^d}{(2(\lambda - (\lambda))^{\frac{1}{k}} + 1)^d} \\ &=
\tau^{d/k} \frac{\left(2\left( 1 - \frac{(\tau \lambda)}{\tau\lambda}\right)^{1/k} + \frac{1}{|\tau\lambda|^{1/k}}\right)^d}{\left(2\left( 1 - \frac{(\lambda)}{\lambda}\right)^{1/k} + \frac{1}{\lambda^{1/k}}\right)^d}. 
\end{split}\end{equation}
Taking the limit over $\lambda$ first and then over $\tau$ we see that
$$
\lim_{\tau \rightarrow 1} \lim_{\lambda \rightarrow \infty} \frac{\phi_V(\tau\lambda)}{\phi_V(\lambda)} = 1.
$$
Lemma \ref{lem1} implies  that 
$$
\int_0^\infty \frac{\lambda^m}{(\lambda + u)^m} ~ d\phi_H(u) /  
\int_0^\infty \frac{\lambda^m}{(\lambda + u)^m} ~ d\phi_V(u) \rightarrow 1 , ~ \mathrm{as} ~ \lambda \rightarrow \infty.
$$

(2) The distribution function 
This in turn implies, by Theorem \ref{gp} of Grishin-Poedintseva, that
\begin{equation}\label{eqn-ratio}
\phi_H(\lambda) / \phi_V(\lambda) \rightarrow 1 , ~ \mathrm{as} ~ \lambda \rightarrow \infty.
\end{equation}
Then (2) follows. 

(3) This follows from (2) and equation (\ref{eqn6}).

(4) Using the asymptotics (3), bounding the terms in the ratio 
$\frac{Tr(\pi_{\mu+r})}{Tr(\pi_\lambda)}$ and keeping the possibility that $r$ also goes to
infinity we get this estimate.\qed

This corollary implies that $\phi_H$ is also a multiplicatively continuous function 
from the following Lemma.
\begin{lemma}\label{lem3}
The function $\phi_H$ considered above is multiplicatively continuous at infinity.
\end{lemma}
\pf We will show that if $\varphi, \chi$ are two distribution functions satisfying
$$
\lim_{r \rightarrow \infty} \frac{\varphi(r)}{\chi(r)} = 1,
$$
then $\varphi$ is multiplicatively continuous whenever $\chi$ is.
Clearly 
$$
\lim_{r \rightarrow \infty} \lim_{\tau \rightarrow 1} \frac{\varphi(\tau r)}{\varphi(r)} = 
\lim_{r \rightarrow \infty} \frac{\varphi(r)}{\varphi(r)} = 1.
$$
Now consider
\begin{equation}\begin{split}
\lim_{\tau \rightarrow 1} \lim_{r \rightarrow \infty} \frac{\varphi(\tau r))}{\varphi(r)} &=
\lim_{\tau \rightarrow 1} \lim_{r \rightarrow \infty} \frac{\frac{\varphi(\tau r))}{\chi(\tau r)}}{\frac{\varphi(r)}{\chi(r)}} \\
& = \lim_{\tau \rightarrow 1} 1 = 1,  
\end{split}\end{equation}
where in the last step we used the assumption on $\phi/\chi$ and the fact that
$\chi$ is multiplicatively continuous.  Since $\phi_V$ is multiplicatively
continuous, the above result together with equation (\ref{eqn-ratio}) now shows that
$\phi_H$ is multiplicatively continuous. \qed

\begin{lemma}\label{lem2}
Suppose $B$ is a bounded positive operator on 
$\ell^2(\ZZ^d)$, then, for $m\in \NN$ for which $\mvlam$ (and hence $\mhlam$)
is trace class, we have,
$$
\vline ~ \frac{Tr(B(H + \lambda)^{-m})}{Tr(B(V+\lambda)^{-m})} - 1 \vline \rightarrow 0
$$
as $\lambda \rightarrow \infty$.
\end{lemma}
\pf The proof is similar to that in the above lemma, except that we have to
do a bit more of algebra in handling the error term, namely, using equation (\ref{expand}) we write 
\begin{equation}\begin{split}
Tr(B\mhlam)  & = Tr(B\mvlam)  \\ & + Tr\left(B \hmvlam \left( (1+K_\lambda)^{-m} - 1)  \right) \hmvlam\right).
\end{split}\end{equation}
we set $W_\lambda = \hmvlam B \hmvlam $ which is a positive trace class, so we rewrite 
the error term as 
\begin{equation}
\begin{split}
& \vline Tr\left(B \hmvlam \left( (1+K_\lambda)^{-m} - 1)  \right) \hmvlam\right)\vline 
\\ & \leq \vline Tr\left(\hmvlam B \hmvlam \left( (1+K_\lambda)^{-m} - 1)  \right) \right)\vline
\\ & \vline Tr\left(W_\lambda^\half \left( (1+K_\lambda)^{-m} - 1) W_\lambda^\half  \right) \right)\vline
\\ & \leq Tr(W_\lambda) \|\left( (1+K_\lambda)^{-m} - 1\right)\| \leq Tr(B \mvlam) m \|\Delta\| \|\vlam\|^m.
\end{split}
\end{equation}
The rest of the proof is as in the Lemma \ref{lem1} using the above
estimate. \qed

\begin{prop}\label{prop1}
Consider $V, H$ as in equations (\ref{eqn0}, \ref{eqn1}).  Then for any bounded positive
operator $B$ and $m \in \NN$ be such that $\mvlam$ is trace class.  
Then we have 
\begin{enumerate}
\item[(i)] The following equality is valid in the sense that if one of the limits exists then the
other also does and the limits are the same.
$$
\lim_{\lambda \rightarrow \infty}  \frac{Tr(B\mhlam)}{Tr(\mhlam)} = 
\lim_{\lambda \rightarrow \infty}  \frac{Tr(B\mvlam)}{Tr(\mvlam)}.  
$$

\item[(ii)] If in addition $B$ comes from an operator of multiplication by a function 
$b$ on $L^2(\TT^d)$, then the limits in (i) exist.
\end{enumerate}
\end{prop}
\pf (i) For each $\lambda$ we have the equality  
$$
\frac{\left(\frac{Tr(B(H + \lambda)^{-m})}{Tr(B(V+\lambda)^{-m})}\right) }{\left(\frac{Tr((H + \lambda)^{-m})}{Tr((V+\lambda)^{-m})}\right) }   = \frac{\left(\frac{Tr(B(H + \lambda)^{-m})}{Tr((H+\lambda)^{-m})}\right) } {\left(\frac{Tr(B(V + \lambda)^{-m})}{Tr((V+\lambda)^{-m})}\right) }.
$$
Lemma \ref{lem1} and Lemma \ref{lem2} imply that the left hand side has limit $1$, hence 
the right hand side limit exists and equals $1$. Therefore if either  
the numerator or the denominator in the fraction in the right hand side has a limit, 
then the other also has a limit and they both agree which implies the proposition. 

(ii) In the case when $B$ comes from an operator of multiplication by a function $b$ on
$L^2(\TT^d)$ we have, using the Fourier series to compute $\langle\delta_n, B \delta_n\rangle$ 
for any $n$,
\begin{equation}\begin{split}
Tr(B\mvlam) &= \sum_{n \in \ZZ^d } \langle \delta_n, B \delta_n \rangle (V(n)+\lambda)^{-m}
\\ &=\left(\frac{1}{(2\pi)^d} \int b(x) dx \right) \sum_{n \in \ZZ^d} (V(n)+\lambda)^{m}
\\ &=\left(\frac{1}{(2\pi)^d} \int b(x) dx \right) Tr(\mvlam). 
\end{split}\end{equation}
Therefore we have for each $\lambda$, 
$$
\frac{Tr(B\mvlam)}{Tr(\mvlam)} = \left(\frac{1}{(2\pi)^d} \int b(x) dx \right),
$$
so the limit of the left hand side as $\lambda$ goes to infinity exists.
\qed

Given these general results, we now use the theorems of Laptev-Safarov (Theorem \ref{ls})  and 
the Tauberian theorem of Grishin-Poedintseva (Theorem \ref{gp}) to pass onto  
measures associated with the self adjoint operators $H$ and $V$.  

{\noindent \bf Proof of Theorem \ref{szego}: }

Since $B$ is bounded self-adjoint $\|B\| \in \sigma(B)$, since $B$ is also positive, 
from the definition of $K$ it is clear that $[0, \|B\|] = [0, 1]\|B\| \subset K$. 
Since $\|\pi_\lambda B \pi_\lambda\| \leq \|B\|$, it is also clear
that for each $\lambda$, $\sigma(\pi_\lambda B \pi_\lambda) \subset K$.  Also since $K$
is compact by definition, the elements of $C(K)$ can be uniformly approximated by
those from $W_\infty^2(K)$ (the space of all twice continuously differentiable functions
equipped with the norm $\||f\|| = \sum_{j=0}^2 |f^{(j)}|_\infty$) by Stone-Weierstrass, since the latter 
contains polynomials. 
Therefore it is enough to prove the theorem for $f \in W_\infty^2(K)$. 

We set $N_r(\lambda) = \sup_{\mu \leq \lambda} (Tr(\pi_{\mu +r}) - Tr(\pi_\mu) ), r>0$.  Then
using the theorem \ref{ls}  we get
\begin{equation}\begin{split}
& \vline ~ \frac{Tr\left( f(\pi_\lambda B \pi_\lambda)\right) }{Tr(\pi_\lambda)} - \frac{Tr\left(\pi_\lambda f(B )\pi_\lambda\right) }{Tr(\pi_\lambda)} ~ \vline \\ & \leq 
\frac{1}{Tr(\pi_\lambda) } \half \|f^{''} \| N_r({\lambda}) \left(\|\pi_{\lambda -r }B\|^2 + \frac{\pi^2 \lambda^{2\kappa}}{6 r^2} \|H^{-\kappa} \pi_{\lambda -r } [H, B]\|^2\right) \\
& \leq  
C \frac{N_{\lambda^\kappa}(\lambda)}{Tr(\pi_\lambda)} \leq C\frac{[\lambda^{\kappa} ]}{[\lambda]} 
\end{split}\end{equation}
For getting the last estimate, we take $r = \lambda^\kappa$, then by the assumptions on 
$B$ the term in the parenthesis on the right hand side in the penultimate estimate above is bounded, 
so using Corollary \ref{cor1}(3), we get the last bound. 
The last term clearly goes to zero as $\lambda \rightarrow \infty$ 
so it is enough to show 
\begin{equation}
\vline ~ \frac{Tr\left(\pi_\lambda f(B)\pi_\lambda\right) }{Tr(\pi_\lambda)} - \frac{1}{(2\pi)^d}\int  f(b(x)) ~ dx\vline ~ \rightarrow 0 , ~~ \mathrm{as} ~~ \lambda \rightarrow \infty
\end{equation}
Let $\pi_{V,\lambda}$ denote the spectral projection $E_V((0, \lambda])$, then
$$
Tr(\pi_{V, \lambda}f(B) \pi_{V, \lambda}) = \sum_{V(n) \leq \lambda} 
\langle \delta_n, f(B) \delta_n\rangle, ~ \mathrm{and} ~ Tr(\pi_{V,\lambda}) = \sum_{V(n)\leq \lambda} 1.
$$
Since, under the Fourier series the basis vectors $|\delta_n\rangle$ go over to
the basis $e^{in\cdot x}$ in $L^2(\TT^d)$ and $B$ is an operator of multiplication
by a bounded positive function $b(x)$ there, we see that 
$$
\langle \delta_n, f(B) \delta_n\rangle = \frac{1}{(2\pi)^d}\int_{\TT^d} f(b(x)) dx.
$$
Therefore, for each $\lambda$,
$$
\frac{Tr(\pi_{V,\lambda} f(B) \pi_{V,\lambda}) }{Tr(\pi_{V,\lambda})} = \frac{1}{(2\pi)^d}\int_{\TT^d} f(b(x)) dx.
$$

Therefore it is enough to show
that  
\begin{equation}\label{eqn5}
\lim_{\lambda \rightarrow \infty} \frac{Tr(\pi_\lambda f(B) \pi_\lambda) }{Tr(\pi_\lambda)} = \lim_{\lambda \rightarrow \infty} \frac{Tr(\pi_{V,\lambda} f(B) \pi_{V,\lambda}) }{Tr(\pi_{V,\lambda})},
\end{equation}
to prove the theorem.  Adding a constant to the function $f$ does not matter in the above, 
so we can assume without loss of generality that the function $f$ is positive, 
so $f(B)$ is a positive operator and hence $f(b(x))$ is a positive function on $\TT^d$.

Now recall the definition of $\phi_H, \phi_V$ and we set
\begin{equation}\begin{split}
\phi_{H,f}(\lambda) &= Tr(\pi_\lambda f(B) \pi_\lambda) = Tr(f(B)^\half \pi_\lambda f(B)^\half), 
 \\
\phi_{V,f}(\lambda) &= Tr(\pi_{V,\lambda} f(B) \pi_{V,\lambda}) = Tr(f(B)^\half \pi_{V,\lambda} f(B)^\half), 
 \\
\end{split}\end{equation}
In this notation, the claim in the equation is nothing but the convergence
$$
\lim_{\lambda \rightarrow \infty} \frac{\phi_{H,f}(\lambda)}{\phi_H(\lambda)} = \lim_{\lambda \rightarrow \infty} \frac{\phi_{V,f}(\lambda)}{\phi_H(\lambda)}. 
$$

This convergence follows if we show, by theorem \ref{gp},
$$
\lim_{r \rightarrow \infty} \frac{\int \frac{\phi_{H,f}(u)}{(1 + \frac{u}{r})^{m+1}} ~ du}{\int \frac{\phi_H(u)}{(1 + \frac{u}{r})^{m+1}} ~ du} = \lim_{r \rightarrow \infty} 
\frac{\int \frac{\phi_{V,f}(u)}{(1 + \frac{u}{r})^{m+1}}~ du}{\int \frac{\phi_V(u)}{(1 + \frac{u}{r})^{m+1}}~ du} 
$$ 
This equality is precisely the content of Proposition \ref{prop1} (where $B$ is replaced by $f(B)$),
after an integration by parts performed in all the integrals above and
using the spectral theorem respectively for $H$ and $V$ to rewrite the integrals as traces.  \qed

{\noindent \bf Proof of Theorem \ref{szego2}: }

The proof is as in the proof of theorem \ref{szego} till equation (\ref{eqn5}), 
the difference comes in 
computing the limit of the right hand side in equation (\ref{eqn5}).  Hence it is enough to 
compute the limits
$$
\lim_{\lambda \rightarrow \infty} \frac{Tr(\pi_{V,\lambda} f(B) \pi_{V,\lambda}) }{Tr(\pi_{V,\lambda})} = \lim_{\lambda \rightarrow \infty}\frac{ \sum_{n : V(n) \leq \lambda } \langle \delta_n, f(B) \delta_n \rangle}{\sum_{V(n) \leq \lambda} 1 } .
$$
when $B$ comes from a symbol $b(x,n)$ and $f$ continuous.  The operator $B$ being
bounded self-adjoint, continuity of $f$ implies that $f(B)$
is approximated in norm by polynomial functions of $B$, by using the spectral theorem 
and the Stone-Weierstrass theorem together, which approximation is uniform in
$\lambda$ in the above limits, hence by an $\epsilon/3$ argument, it is enough to consider
$f$ to be a polynomial of a fixed degree and further by linearity of the limits it is 
enough to take $f$ to be a monomial in $B$.

Therefore we need to show that for any $k \in \NN$, 
$$
\lim_{\lambda \rightarrow \infty}\frac{ \sum_{n : V(n) \leq \lambda } \langle \delta_n, B^k \delta_n \rangle}{\sum_{V(n) \leq \lambda} 1 } = 
\lim_{\lambda \rightarrow \infty}\frac{ \sum_{n : V(n) \leq \lambda } \frac{1}{(2\pi)^d} \int (b(x,n))^k dx }{\sum_{V(n) \leq \lambda} 1 },
$$
when $B$ comes from a symbol $b \in S_{1,0}(\TT^d\times \ZZ^d)$ to prove the theorem.

Therefore a simple computation shows that firstly
if $\sigma_{B^k}(x,n)$ is the symbol associated with $B^k$ then,  
$$
\langle \delta_n, B^k \delta_n\rangle = \langle U\delta_n, (UBU^*)^k U\delta_n\rangle 
= \frac{1}{(2\pi)^d} \int \sigma_{B^k}(x,n) ~ dx.
$$

Secondly the symbol $\sigma_{B^k}(x,n)$ is given in terms of $b(x,n)$ asymptotically in $n$
using the above relation applied $k$ times on $b$.  Using Lemma \ref{lem4} below we see that
$$
\langle \delta_n, B^k \delta_n\rangle = \frac{1}{(2\pi)^d}\int {b(x,n)}^k ~ dx + 
\frac{1}{(2\pi)^d}\int {E_k(x,n)} ~ dx
$$ 
with $\sup_{x} | E_k(x,n)| \rightarrow 0$ as $|n| \rightarrow 0$. 

Then computing the limits and using the fact that if a sequence $r_n$ goes to zero as $|n|$ goes
to infinity then $\left(\sum_{V(n) \leq \lambda} r_n\right) /\sum_{V(n) \leq \lambda}$ goes to zero as 
$\lambda$ goes to infinity, by an $\epsilon/3$ argument.  Therefore using the properties
of $E_k(x,n)$ stated above together with the Lebesgue dominated convergence theorem, we see that
\begin{equation}
\begin{split}
\lim_{\lambda \rightarrow \infty}\frac{ \sum_{n : V(n) \leq \lambda } \langle \delta_n, B^k \delta_n \rangle}{\sum_{V(n) \leq \lambda} 1 } & = \lim_{\lambda \rightarrow \infty}\frac{ \sum_{n : V(n) \leq \lambda } \frac{1}{(2\pi)^d} \int (b(x,n))^k ~ dx }{\sum_{V(n) \leq \lambda} 1 } \\
& + \lim_{\lambda \rightarrow \infty}\frac{ \sum_{n : V(n) \leq \lambda } \frac{1}{(2\pi)^d} \int E_k(x,n) ~ dx}{\sum_{V(n) \leq \lambda} 1 } \\
& = \lim_{\lambda \rightarrow \infty}\frac{ \sum_{n : V(n) \leq \lambda } \frac{1}{(2\pi)^d} \int (b(x,n))^k ~ dx }{\sum_{V(n) \leq \lambda} 1 }.
\end{split}
\end{equation}
This proves the theorem. \qed

From theorem 4.2 \cite{mic}, we see that if $A, B$ are
linear operators in $L^2(\TT^d)$ associates with symbols
$a, b \in S_{1,0}^0(\TT^d\times\ZZ^d)$, then $AB$ also comes from a symbol $\sigma(x,n) \in S_{1,0}^0(\TT^d\times \ZZ^d)$ and
\begin{equation}\label{compose}
\sigma(x,n) \approx \sum_{\alpha \geq 0} \frac{1}{\alpha !} (\Delta_m^\alpha a(x,n)) D_x^{(\alpha)} b(x,n)
\end{equation}
where $\approx$ means asymptotic in $n$.  

\begin{lemma}\label{lem4}
Consider a symbol $a(x,n) \in S_{1,0, \infty}(\TT^d\times\ZZ^d)$ and let $A$ be the pseudo
difference operator on $L^2(\TT^d)$ associated with it.
Then  for any $k \in \NN$, the symbol 
$a_{k}(x,n)$ of the operator $A^k$ has the asymptotic behaviour 
$$
a_{k}(x,n) \approx (a(x,n))^k  + E_{k}(x,n),  
$$ 
with $E_k(x,n) \in S_{1,0,\infty}^{-1}$. 
\end{lemma}
\pf We will prove this by induction, since for $k=1$ this is trivial with $E_{1}(x,n) = 0$.
We assume that the Lemma is valid for $a_{k-1}(x,n)$, so we assume that
$$
a_{k-1}(x,n) \approx (a(x,n))^{k-1}  + E_{k-1}(x,n), ~ \mathrm{with} ~ E_{k-1} \in S_{1,0,\infty}^{-1}. 
$$
We use the composition rule in equation (\ref{compose}) to get 
\begin{equation}\label{comp2}
\begin{split}
a_{k}(x,n) &\approx \sum_{\alpha \geq 0} \frac{1}{\alpha !} (\Delta_n^\alpha a(x,n)) D_x^{(\alpha)}a_{k-1}(x,n) \\
&\approx \sum_{\alpha \geq 0}\frac{1}{\alpha !}  (\Delta_n^\alpha a(x,n)) D_x^{(\alpha)}(a(x,n))^{k-1} 
+ \sum_{\alpha \geq 0}\frac{1}{\alpha !} (\Delta_n^\alpha a(x,n)) D_x^{(\alpha)}E_{k-1}(x,n) \\
& \approx (a(x,n))^k + T_1(x,n) + T_2(x,n),
\end{split}
\end{equation} 
where
\begin{equation}\begin{split}
T_1(x,n) &= \sum_{|\alpha| \geq 1}\frac{1}{\alpha !}  (\Delta_n^\alpha a(x,n)) D_x^{(\alpha)}(a(x,n))^{k-1}, \\
T_2(x,n) &=
\sum_{\alpha \geq 0}\frac{1}{\alpha !} (\Delta_n^\alpha a(x,n)) D_x^{(\alpha)}E_{k-1}(x,n).
\end{split}\end{equation}
We set $E_{k}(x,n) = T_1(x,n) + T_2(x,n)$ and prove its properties stated in the assumption.
We recall the relation 
$$
\Delta_{n}^\alpha \sigma(n) = \sum_{\beta\leq \alpha} (-1)^{|\alpha - \beta|} \left(\begin{matrix}\alpha \\ \beta\end{matrix}\right) \sigma(n+\beta).
$$
from Proposition 3.1 in \cite{mic}.  Using this and the facts that $a \in S_{1,0, \infty}$,
$E_{k-1} \in S_{1,0, \infty}^{-1}$ we estimate
$$
|\Delta_n^\alpha a(x,n)| \leq C 2^{|\alpha|}, |D_x^\alpha E_{k-1}(x,n)| \leq C \langle n \rangle^{-1},
$$ 
so that each of the terms in the sum defining $T_2(x,n) \in S_{1,0,\infty}^{-1}$ and 
$$
|T_2(x,n)| \leq C \langle n \rangle^{-1}.
$$
so that $T_2(x,n) \in S_{1,0,\infty}^{-1}$.  To estimate $T_1(x,n)$ we define   
multi-indices $\alpha^{(j)}$ to be $\alpha^{(j)}_r = \delta_{rj}, r=1,\dots, d$,
and split $T_1$ as 
$$
T_1(x,n) = \sum_{j=1}^d \sum_{\alpha \geq 1,  \alpha_j \geq 1}\frac{1}{\alpha !}  (\Delta_n^{\alpha - \alpha^{(j)}} \Delta_n^{\alpha^{(j)}}a(x,n)) D_x^{(\alpha)}(a(x,n))^{k-1}.
$$
Then clearly $\Delta_n^{\alpha^{(j)}}a(x,n) \in S_{1,0,\infty}^{-1}$.
If 
$a,b \in S_{1,0}^0$,then $ab \in S_{1,0}^0$, however the
same is not true for $S_{1,0,\infty}^0$ in view of Leibniz rule
for derivatives. Therefore
using the property of $a \in S_{1,0, \infty}$, we estimate
$$
|(\Delta_n^{\alpha - \alpha^{(j)}} \Delta_n^{\alpha^{(j)}}a(x,n))| \leq C 2^{|\alpha|} \langle n \rangle^{-1}, ~ \frac{1}{\alpha !} |D_x^{(\alpha)}(a(x,n))^{k-1}| \leq C^{k-1}
\prod_{j=1}^d \theta_j,
$$
where 
$$
\theta_j = \begin{cases} 1, ~~ \alpha_j \leq k-1, \\ \frac{1}{(\alpha_j - k +1)!}, ~~ \alpha_j - k + 1 > 0 \end{cases} , ~~ j=1, \dots , d,
$$
Then arguing as done for the term $T_2(x,n)$ we see that $T_1(x,n) \in S_{1,0,\infty}^{-1}$
and 
$$
|T_1(x,n)| \leq C \langle n \rangle^{-1},
$$
proving the lemma. \qed

{\noindent \bf Acknowledgement:} We thank M N Namboodiri for introducing us to the Szeg\"{o} type theorems and
referees for numerous comments on earlier versions of this paper that made us simplify
and clarify the proofs here.

\section{Appendix}

In this appendix we collect two theorems we use in our paper for the reader's convenience.

The first one is a Tauberian theorem of Grishin-Poenditsheva from \cite{gri}. 

\begin{defn}
 Let $\phi$ be a positive function on the half line $[0,\infty)$. Let 
$$
S = \{ \alpha : \exists M,R ~ \mathrm{with} ~ \phi(tr) \leq M t^\alpha, ~ \mathrm{for ~ all ~ }  t \geq 1, r \geq R \} 
$$
and
$$
G = \{ \alpha : \exists M,R ~ \mathrm{with} ~ \phi(tr) \geq M t^\alpha, ~ \mathrm{for ~ all ~ }  t \geq 1, r \geq R \}
$$
Then {\bf $\alpha(\phi)=\inf S$} and  
{\bf $\beta(\phi)=\sup G$}  are called the {\bf upper and lower Matushevskaya index}
 of $\phi$ respectively.

\end{defn}

\begin{thm}(\cite{gri},Theorem 2)\label{r}

Let $m>-1$. Assume that $\varphi$ is positive measurable function on $[0,\infty)$ that 
does not vanish identically in any neighbourhood of infinity. Let 
$\Phi(r)=\displaystyle\int_0^\infty \dfrac{\varphi(rt)}{(1+t)^{m+1}}dt$
be finite. 
Then the functions $\varphi$ and $\Phi$ have same growth at infinity if and only if 
$\beta(\varphi)>-1$ and $\alpha(\varphi)<m$.
\end{thm}

\begin{defn}
A function $\varphi$ is said to be multiplicatively continuous at infinity if it satisfies
$$
\lim_{\substack{r\rightarrow \infty\\\tau\rightarrow 1}}\dfrac{\varphi(\tau r)}{\varphi(r)}=1 ~~ \mathrm{and} ~~  
\lim_{\substack{\tau \rightarrow \infty\\ r \rightarrow 1}}\dfrac{\varphi(\tau r)}{\varphi(r)}=1 .
$$
\end{defn}

\begin{thm}(\cite{gri},Theorem 8)\label{gp}
 Let $\varphi$ and $\psi$ be positive functions on $[0,\infty)$ satisfying the following conditions:
\begin{enumerate}
 \item the functions $\varphi$ and $\psi$ do not vanish identically in any neighbourhood of infinity;
\item the function $\varphi$ is multiplicatively continuous at infinity and $\beta(\varphi)>-1$;
\item the function $\psi$ is increasing;
\item at least one of the inequalities $\alpha(\varphi)<m$ and $\alpha(\psi)<m$ holds, where $m>-1$;
\item the functions 
$$
\Phi(r)=\int_0^\infty \dfrac{\varphi(ru)}{(1+u)^{m+1}}du ~~and~~\Psi(r)=\int_0^\infty \dfrac{\psi(ru)}{(1+u)^{m+1}}du
$$
are finite and $\displaystyle\lim_{r\to \infty}\dfrac{\Psi(r)}{\Phi(r)}=1$ then $\displaystyle\lim_{r\to \infty}\dfrac{\psi(r)}{\varphi(r)}=1$.
\end{enumerate}
 \end{thm}

The above theorem derives asymptotic behaviour of $\varphi,\psi$ from the 
asymptotic behaviour of $\phi,\Psi$ by assuming additional conditions on 
$\varphi$ and $\psi$.

The next theorem is Theorem 1.2 of  Laptev - Safarov \cite{LapSaff}.

Let $\hh$ be a separable Hilbert space and $B$ a self adjoint operator, not necessarily
bounded.  Suppose that 
$f \in W_\infty^2(K)$ where $K = \bigcup_{0 \leq t \leq 1} t\sigma(B) \subset \RR$ and
suppose $\pi_\lambda = E_H((0, \lambda))$.

With these conditions the theorem is :

(Take  $A = H, P_\lambda = \pi_\lambda$ in Theorem 1.6 \cite{LapSaff} to get the
following restatement of their theorem, so as to be consistent with our notation.)

\begin{thm}(Theorem 1.6, \cite{LapSaff})\label{ls} Let  $H >0$ and $rank(\pi_\lambda) < \infty$.
Then for all $f \in W_\infty^2(K)$ and for all $\lambda >0, r >0, \kappa \geq 0$ we have
\begin{equation}\begin{split}
&\vline ~ Tr(\pi_\lambda f(B) \pi_\lambda - Tr (\pi_\lambda f(\pi_\lambda B \pi_\lambda) \pi_\lambda) ~ \vline \\
&\leq \half \|f^{''}\| N_r(\lambda) \left(\|\pi_{\lambda -r }B\|^2 + \frac{\pi^2 \lambda^{2\kappa}}{6 r^2} \|H^{-\kappa} \pi_{\lambda -r } [H, B]\|^2\right).
\end{split}\end{equation}
\end{thm}

\end{document}